\documentclass[twocolumn,showpacs]{revtex4}
\usepackage{epsfig}
\usepackage{graphicx}%
\usepackage{dcolumn}
\usepackage{amsmath}

\makeatletter
\def\btt#1{\texttt{\@backslashchar#1}}%
\DeclareRobustCommand\bblash{\btt{\@backslashchar}}%
\makeatother

\begin{document}

\title{ Cosmological perturbations and noncommutative tachyon inflation}

\author{Dao-jun Liu}
\author{Xin-zhou Li}\email{kychz@shnu.edu.cn}
\affiliation{Shanghai United Center for Astrophysics(SUCA),
Shanghai Normal University, 100 Guilin Road, Shanghai 200234,China
}%

\date{\today}

\begin{abstract}
ABSTRACT: The motivation for studying the rolling tachyon and
non-commutative inflation comes from string theory. In the tachyon
inflation scenario, metric perturbations are created by tachyon
field fluctuations during inflation. We drive the exact mode
equation for scalar perturbation of the metric and investigate the
cosmological perturbations in the commutative and non-commutative
inflationary spacetime driven by the tachyon field which have a
Born-Infeld Lagrangian.
\end{abstract}

 \maketitle

\section{Introduction}
The cosmological parameters and the  properties of inflationary
models are tightly constraint by the recent result from Wilkinson
Microwave Anisotropy Probe (WMAP)\cite{WMAP} and other earlier
observations. The standard inflationary $\Lambda$CDM model
provides a good fit to the observed cosmic microwave background
(CMB) anisotropies. The first-year results of WMAP also bring us
something intriguing, some analyses
\cite{Bridle,Mukherjee,Gaztanaga,Kesden} show that the new data of
CMB suggesting an anomalously low quadrupole and octupole and a
larger running of the spectral index of the power spectrum than
that predicted by standard single scalar field inflation models
satisfying the slow roll conditions.

 One typically considers  an inflationary phase driven by the potential of the inflation ,
 whose dynamics is determined by a canonical scalar action. Recently, pioneered by Sen
\cite{Sen}, the study of non-BPS objects such as non-BPS branes,
brane-antibrane configurations or space-like branes has attracted
physical interests in string theory. Sen showed that classical
decay of unstable D-brane in string theories produces pressureless
gas with non-zero energy density. Gibbons took into account the
gravitational coupling by adding an Einstein-Hilbert term to the
effective action of the tachyon on a brane, and initiated a study
of "tachyon cosmology" \cite{Gibbons}. This provides a rich gamut
of possibilities in the context of cosmology, including slow-roll
inflation \cite{sami}. As an inflationary mechanism, tachyon
condensation has been criticized by some authors \cite{Kofman}.
Their reason is that for string theory motivated values of the
parameters in potential $V(T)$, there is an incompatibility
between the slow-roll condition and COBE normalization of
fluctuations. However, the potential can be found for which this
issue may be circumvented \cite{Bento}. Thus, one can take a
phenomenological approach and study the inflationary predictions
by the tachyon field. Tachyon inflation leads to a deviation in
one of the second order consistency relations, and its predictions
are typically characteristic of small field or chaotic inflation
\cite{Steer}. All the typical tachyon models predict a negative
and very small running of the scalar spectral index, and they
consistently lie within the $1\sigma$ contour of the data set.
However, the regime of blue scalar spectral index and large
gravitational waves is not explored by these model.

On the other hand, it is well known that during the period of
inflation, the classical gravitational theory, general relativity,
might break down due to the very high energies at that time and
the correction from string theory may take effect. In the
nonperturbative string/M theory, any physical process at the very
short distance takes an uncertainty relation, called stringy
spacetime uncertainty relation (SSUR),
\begin{equation}
\Delta t_p \Delta x_p \geq l_s^2,
\end{equation}
where $t_p$ and $x_p$ are the physical time and space, $l_s$ is
the string length scale. It is suggested that the SSUR is a
universal property for strings as well as D-branes \cite{Yoneya}.
 Unfortunately, we now have no ideas to derive
cosmology directly from string/M theory. Brandenberger and Ho
\cite{Brandenberger} have proposed a variation of spacetime
non-commutative field theory to realize the stringy spacetime
uncertainty relation without breaking any of the global symmetries
of the homogeneous isotropic universe. If inflation is affected by
physics at a scale close to string scale, one expects that
spacetime uncertainty must leave vestiges in the CMB power
spectrum \cite{Huang,Tsujikawa,Huang2,Huang3}. It is found that
the modification from the non-commutative spacetime or SSUR
delayed the cosmological perturbation mode crosses the Hubble
horizon for a smaller Hubble constant and thus suppress the
fluctuation which implies that the running of the spectral index
is larger than the one in the commutative case.

The motivation for studying the tachyon inflation comes from type
II string theory. It is not unique, but has its counterpart. The
non-commutative inflation which takes into account some effects of
the spacetime uncertainty principle motivated by ideas from string
theory. Therefore, in this paper, we investigate cosmological
perturbations of the metric during
 the tachyon inflation in non-commutative
 spacetime. Using the "Mukhanov variable" $z$, after a prolix but
 straightforward calculation, we show the exact mode equation for
 the scalar perturbation of metric. In the tachyon inflation
 scenario, we conform that the non-commutative spacetime effects
 always suppress the power spacetime of both the scalar and
 tensor perturbations, and may provide a large enough running of
 the spectral index to fit the WMAP data.

\section{Hamilton-Jacobi equation of Tachyon inflation}
The flat FRW line element is given by:
\begin{eqnarray}\label{bgMetric}
ds^2=& &dt^2-a^2(t)(dx^2+dy^2+dz^2)\nonumber\\
    =& &a^2(\tau)[d\tau^2-(dx^2+dy^2+dz^2)]
\end{eqnarray}
\noindent where $\tau$ is the conformal time, with $dt=ad\tau$.
The Lagrangian density of a rolling tachyon is
\begin{equation}\label{lagrangian}
L=\sqrt{-g}\left(\frac{R}{2\kappa}-V(T)\sqrt{1-g^{\mu\nu}\partial_\mu
T\partial_\nu T}\right)
\end{equation}
\noindent where $\kappa=8\pi G=M_p^{-2}$. For a spatially
homogenous tachyon field $T$, we have the equation of motion
\begin{equation}\label{ddT}
\ddot{T}+3H\dot{T}\left(1-\dot{T}^2\right)+\frac{V'}{V}\left(1-\dot{T}^2\right)=0
\end{equation}
\noindent which is equivalent to the entropy conservation
equation. Here, the Hubble parameter $H$ is defined as $H\equiv
(\frac{\dot{a}}{a})$, and $V'=dV/dT$. If the stress-energy of the
universe is dominated by the tachyon field $T$, the Einstein field
equations for the evolution of the background metric,
$G_{\mu\nu}=\kappa T_{\mu\nu}$, can be written as
\begin{equation}\label{H2}
H^2=\left(\frac{\dot{a}}{a}\right)^2=\frac{\kappa}{3}\frac{V(T)}{\sqrt{1-\dot{T}^2}}
\end{equation}
\noindent and
\begin{equation}\label{dda}
\frac{\ddot{a}}{a}=H^2+\dot{H}=\frac{\kappa}{3}\frac{V(T)}{\sqrt{1-\dot{T}^2}}\left(1-\frac{3}{2}\dot{T}^2\right)
\end{equation}
\noindent Eqs.(\ref{ddT})-(\ref{dda}) form a coupled set of
evolution equations of the universe. The fundamental quantities to
be calculated are $T(t)$ and $a(t)$, and the potential $V(T)$ is
given when the model is specified. The period of accelerated
expansion corresponds to $\dot{T}^2<\frac{2}{3}$ and decelerate
otherwise. In the limit case $\dot{T}=0$, there is no difference
in meaning of the expansion of universe between tachyon inflation
and ordinary inflation driven by inflaton. However, the case of
$\dot{T}\neq 0$ forms a sharp contrast. Although the formulas of
tachyon inflation are correspond to those of the inflation driven
by ordinary scalar field, there is obvious difference between them
which can not be neglected. From Eqs.(\ref{ddT})-(\ref{dda}), we
have two first-order equations
\begin{equation}\label{T7}
\dot{T}=-\frac{2}{3}\frac{H'(T)}{H^2(T)}
\end{equation}
\begin{equation}\label{HJE:8}
 [H'(T)]^2-\frac{9}{4}H^4(T)=-\frac{\kappa^2}{4}V^2(T)
\end{equation}
\noindent These equations are wholly equivalent to the
second-order equation of motion (\ref{ddT}).

Analogous to the inflation driven by ordinary scalar field, for
example in Ref.\cite{Kolb}, we need define the "slow-roll"
parameters. In general, there are two ways to define them. One is
that we can take the definitions are independent of field driving
inflation \cite{Steer}, $\epsilon\equiv H_{\ast}/H$ and
$\epsilon_{i+1}=d\ln|\epsilon_i|/dN$ ($i\geq 0$), where $H_{\ast}$
is the Hubble parameter at some chosen time. They are an
advantageous choice to use in order to compare ordinary and
tachyon inflation, though the observables (such as spectral
indices) will no longer be related to the parameters in the same
way. And the other is called tachyonic slow-roll parameters as
follows
\begin{eqnarray}\label{epsilon:9}
\epsilon(T)&\equiv&
\frac{2}{3}\bigg(\frac{H'(T)}{H^2(T)}\bigg)^2,\\
\eta(T)&\equiv&\frac{2}{3}\bigg(\frac{H''(T)}{H^3(T)}\bigg),\label{eta10}\\
\xi(T)&\equiv&
\frac{2}{3}\bigg(\frac{H'(T)H'''(T)}{H^6(T)}\bigg)^{1/2}.\label{xi11}
\end{eqnarray}
Obviously, the definitions of the parameters
Eqs.(\ref{epsilon:9})-(\ref{xi11}) are quite different from those
defined in ordinary inflation. This is very natural for the
Born-Infeld action is sharply different from that of the ordinary
scalar field. In next section these parameters will be
conveniently applied to exact mode equation of tachyon inflation.
In term of $\epsilon(T)$ parameter, Eq.(\ref{HJE:8}) can be
reexpressed as
\begin{equation}\label{vh:12}
H^4(T)[1-\frac{1}{3}\epsilon (T)]=\frac{\kappa^2}{9}V^2(T)
\end{equation}
\noindent which is referred to as the Hamilton-Jacobi equation of
tachyon inflation. Using Eq.(\ref{T7}), we have
\begin{equation}
\epsilon(T)=\frac{3}{2}\dot{T}^2.
\end{equation}
Note that the Hamilton-Jacobi equation has the same form as that
of the ordinary inflaton field only up to first order term in
$\epsilon(T)$. This can be found by comparing the Hamilton-Jacobi
equation for an ordinary scalar field \cite{Kinney} with the one
for the tachyon, Eq.(\ref{vh:12}),
\begin{equation}
H^2\left[1-\frac{1}{3}\epsilon(T)\right]=\frac{\kappa}{3}\frac{(1-\frac{1}{2}\dot{T}^2)V(T)}{\sqrt{1-\dot{T}^2}}=\frac{8\pi
G}{3}V(T)+O(\epsilon^2)
\end{equation}

 The number of e-folds of the inflation produced
when the tachyon field rolls from a particular value $T$ to the
end point $T_e$ is
\begin{equation}
N(T,T_e)\equiv \int^{t_e}_t
H(t)dt=\int^{T_e}_{T}\frac{H}{\dot{T}}dT
\end{equation}
\noindent Therefore, we have
\begin{equation}
a(T)=a_e \exp[-N(T)]
\end{equation}
\noindent where $a_e$ is the value of the scale factor at the end
of inflation. Since after tachyon inflation the dynamics of the
reheating is still unclear, in the following we shall typically
assume a conservative value of e-folds $40\leq N\leq 70$
\cite{Steer}.

Given a non-commutative spacetime that obeys the stringy spacetime
uncertainty relation, the cosmological background will still be
described by the Einstein equations since the background fields
only depend on one spacetime variable \cite{Brandenberger}. Thus,
the formula for the tachyon field that drives the non-commutative
spacetime inflating have the same form as in the ordinary
commutative spacetime. But the equations for the linear
fluctuations should be modified. Brandenberger and Ho
\cite{Brandenberger} argued that the modifications take the form
of an interaction of the fluctuating field with the background
which is nonlocal in time.

\section{the cosmological perturbations in commutative spacetime}

During inflation, quantum fluctuations are stretched on scale
larger than the horizon. There they are frozen until they reenter
the horizon after inflation. Regardless of the field which drives
inflation, a quasi scale invariant spectrum are generated for
large scale perturbations. The most important observational test
of inflation is observation of the Cosmic Microwave Background
(CMB) radiation. Temperature fluctuations in the CMB can be
related to perturbations in the metric at the surface of last
scattering. The metric perturbations are created by tachyon
fluctuations during inflation. In the inflation scenario, quantum
fluctuations on small scales are rapidly red-shifted to scales
much larger than the horizon size. The metric perturbations can be
decomposed according to their spin with respect to a local
rotation of the spatial coordinates on hypersurfaces of constant
time. This leads to two types: scalar, or curvature perturbations,
which couple to the tachyon and form the "seeds" for structure
formation, and tensor, or gravitational wave perturbations, which
do not couple to tachyon. Both scalar and tensor perturbations
contribute to CMB anisotropy.

Considering small fluctuations of the tachyon field, that is
\begin{equation}
T(t,\textbf{x})=T_0(t)+\delta T(t,\textbf{x})
\end{equation}
and one can take the metric of the "perturbed universe" in the
longitudinal gauge as \cite{Mukhanov}
\begin{equation}
ds^2=(1+2\Phi)dt^2-(1-2\Phi)a^2(t)\delta_{ij}dx^idx^j
\end{equation}
where $\Phi$ is the newtonian gravitational potential. The
linearized Einstein equations can be written as
\begin{eqnarray}
\dot{\chi}&=&\frac{a}{H^2}\frac{V(T)\dot{T}^2}{\sqrt{1-\dot{T}^2}}\zeta\label{dchi}\vspace{0.9cm}\\
\dot{\zeta}&=&\frac{H^2}{a^3}\frac{(1-\dot{T}^2)^{3/2}}{V(T)\dot{T}^2}\nabla^2\chi\label{dzeta}
\end{eqnarray}
where the new variables $\chi$ and $\zeta$ are respectively
defined as
\begin{equation}
\chi\equiv \frac{2a}{\kappa^2
H}\Phi,\hspace{1cm}\zeta\equiv\Phi+H\frac{\delta T}{\dot{T}}.
\end{equation}
The intrinsic curvature perturbation of the comoving hypersurfaces
$\zeta$ is gauge invariant. It is not difficult to show that one
can relate the fluctuation of the gravitational potential $\Phi$
to the fluctuation of the tachyon field $\delta T$ on superhorizon
scales. The canonical quantization variable $u$ is defined as
$u\equiv z\zeta$, where
\begin{equation}\label{z19}
z=\frac{a}{H}\sqrt{\frac{V(T)\dot{T}^2}{(1-\dot{T}^2)^{3/2}}},
\end{equation}
is the so-called "Mukhanov variable" \cite{Garriga}. From
(\ref{dchi}) and (\ref{dzeta}), we have
\begin{equation}\label{19}
\left[az^2\left(\frac{u}{z}\right)^{.}\right]^{.}=\frac{z}{a}(1-\dot{T}^2)\nabla^2u.
\end{equation}
Using the conformal time $\tau$ instead of physical time $t$,
after a prolix but straightforward calculation, Eq.(\ref{19}) can
be reduced to
\begin{equation}\label{SPEQ1}
\frac{d^2u}{d\tau^2}-(1-\frac{2}{3}\epsilon)\nabla^2u-\frac{1}{z}\frac{d^2z}{d\tau^2}u=0,
\end{equation}
where
\begin{eqnarray}\label{hh}
\frac{1}{z}\frac{d^2z}{d\tau^2}=2a^2H^2\bigg[1&+&\frac{1}{(1-\frac{2}{3}\epsilon)^2}
\bigg(\frac{5}{2}\epsilon-\frac{3}{2}\eta \nonumber\\
&+&\frac{11}{2}\epsilon^2-4\epsilon\eta+\frac{1}{2}\eta^2+\frac{1}{2}\xi^2\nonumber\\
&+&\frac{4}{3}\epsilon^2\eta+\frac{2}{3}\epsilon\eta^2
-\frac{1}{3}\epsilon\xi^2 \bigg)\bigg].
\end{eqnarray}

As usual, the zeroth order term $2a^2H^2$ ensures that the
spectrum is scale invariant. Expanding the quantity $u$ in Fourier
modes
\begin{equation}
u(\tau,\textbf{x})=\int
\frac{d^3\textbf{k}}{(2\pi)^{3/2}}u_k(\tau)e^{i
\textbf{k}\cdot\textbf{x}},
\end{equation}
the mode function $u_k$ satisfies the following equation
\begin{equation}\label{Eq:uk}
\frac{d^2u_k}{d\tau^2}+\left[(1-\frac{2}{3}\epsilon)k^2-\frac{1}{z}\frac{d^2z}{d\tau^2}\right]u_k=0.
\end{equation}
Therefore, the spectrum of curvature perturbation $P_R(k)$ as
function of wavenumber $k$ could be expressed as
\begin{equation}\label{PRk:17}
P^{1/2}_R(k)=\sqrt{\frac{k^3}{2\pi^2}}\left|\frac{u_k}{z}\right|.
\end{equation}
 \noindent Clearly, the above expression Eq.(\ref{hh}) is
different from that of the ordinary scalar field because the
coupling of curvature perturbations to the stress-energy of
tachyon field is in a very different manner.

The Born-Infeld action is quite different from that of the
ordinary scalar field. Therefore, although the expressions of
tachyon inflation correspond to those of the inflation driven by
the ordinary scalar field, there is obvious difference between
them, which can not be neglected. We find that the usual relation
between the scalar and tensor spectral index is modified.
Therefore, at least in principle, tachyon inflation is
distinguishable from standard inflation. Here, it is worth noting
that some authors \cite{Steer} have obtained the mode equation of
tachyon inflation where the slow-roll approximation is appealed
from the beginning of the reduction, but the expressions given
above are all exact without slow roll approximation. Especially,
at lowest order of slow roll formalism the predictions of ordinary
and tachyon inflation are shown to be the same. Higher order
deviation are present in Eq.(\ref{hh}).

\section{ perturbation spectrum in Non-commutative Tachyon inflation }

The action which reproduces the equation of motion (\ref{SPEQ1})
can be written as
\begin{equation}\label{action1}
S=\frac{1}{2}\int d\tau d^3x
z^2[(\partial_{\tau}\zeta)^{\dag}(\partial_{\tau}\zeta)-c_s^2(\nabla\zeta)^{\dag}(\nabla\zeta)],
\end{equation}
where intrinsic curvature perturbation $\zeta$ and mode function
$u$ have the relationship $z\zeta=u$, and $c_s$ denotes sound
velocity which determined by background tachyon field. In the
momentum space, Eq. (\ref{action1}) can be rewritten as
\begin{equation}\label{action1.5}
S=\frac{1}{2}V_T\int d\tau d^3k
z^2[(\partial_{\tau}\zeta_{-k})(\partial_{\tau}\zeta_k)-c_s^2k^2\zeta_{-k}\zeta_{k}],
\end{equation}
where $V_T$ is the total spatial coordinate volume and $k$ is the
comoving wave number. Following the similar process proposed in
Ref.\cite{Brandenberger}, the SSUR is compatible with an unchanged
homogeneous background, but it leads to changes in the action for
the metric fluctuation. We get a model with non-commutative
modifications
\begin{equation}\label{action2}
S=\frac{1}{2}V_T\int d\tilde{\tau} d^3k
z_k^2[\zeta_{-k}'\zeta_k'-c_s^2k^2\zeta_{-k}\zeta_{k}],
\end{equation}
where $z_k$  is defined as
\begin{equation}
z_k=(\beta_{k}^{+}\beta_k^{-})^{1/4}z,
\end{equation}
in which $z$ is the "Mukhanov variable" defined in Eq.(\ref{z19}),
and
\begin{equation}
\beta_k^{\pm}=\frac{1}{2}[a^{\pm2}({\hat{\tau}}+kl_s^2)+a^{\pm2}({\hat{\tau}}-kl_s^2)].
\end{equation}
Here, the new time variable $\hat{\tau}$ is defined as
$d\hat{\tau}=a^2d\tau$. The primes appeared in the
Eq.(\ref{action2}) denote the derivative with respect to the new
time variable $\tilde{\tau}$, and $\tilde{\tau}$ is related to the
conformal time $\tau$ via
\begin{equation}\label{tau33}
 d\tilde{\tau}=a^2\left(\frac{\beta_k^{-}}{\beta_k^{+}}\right)^{1/2}d\tau.
\end{equation}
Apparently, if the string length scale $l_s$ goes to zero, the
action (\ref{action2}) will reduce to the action (\ref{action1.5})
for the fluctuations in the classical spacetime, which leads to
the equation of motion of perturbations (\ref{Eq:uk}). From the
action (\ref{action2}), the equation of motion of the scalar
perturbations can be written as
\begin{equation}\label{uk''}
u_k''+\left(c_s^2k^2-\frac{z_k''}{z_k}\right)u_k=0,
\end{equation}
where the mode function is defined by $u_k=z_k\zeta_k$ and the
sound velocity $c_s$ satisfies
\begin{equation}
c_s^2=1-\frac{2}{3}\epsilon.
\end{equation}
Let
\begin{equation}
\lambda=\frac{H^2k^2}{a^2M_s^4},
\end{equation}
where $k$ is the comoving wave number of a perturbation mode, and
$M_s=l_s^{-1}$ is the string mass scale. $\lambda$ is a small
dimensionless quantity, because we assume the string mass scale
$M_s$ is very large. Using the slow-roll parameters and
$\frac{1}{z}\frac{d^2 z}{d\tau^2}$ in Eq.(\ref{hh}), we get
\begin{eqnarray}
\frac{z_k''}{z_k}&=&\frac{1}{z}\frac{d^2
z}{d\tau^2}\left[1-2(1+\epsilon)\lambda\right]\nonumber\\
&+&2a^2H^2\lambda
\left[3\epsilon\eta-2\epsilon^2+5\epsilon+1+\frac{3\epsilon(2\epsilon-\eta)(
\eta-\epsilon+1)}{3-2\epsilon}\right],\label{ddzk}
\end{eqnarray}
up to the first order of $\lambda$. Clearly, when $l_s\rightarrow
0$ or $M_s\rightarrow \infty$, the quantity $z_k''/z_k$ and
$\tilde{\tau}$ will be reduced to
 $\frac{1}{z}\frac{d^2 z}{d\tau^2}$ and $\tau$ respectively, and
 then the motion equation (\ref{uk''}) of the mode $\mu_k$ in noncommutative spacetime will
 recover the one in ordinary commutative spacetime (\ref{Eq:uk}).

In the slow-roll approximation, the conformal time $\tau$ can be
expressed approximately by
\begin{equation}
\tau\simeq-\frac{1+\epsilon}{aH}.
\end{equation}
From Eq.(\ref{tau33}), we have
\begin{equation}
\tau\simeq(1-\lambda)\tilde{\tau},
\end{equation}
and then,
\begin{equation}\label{tau39}
\tilde{\tau}\simeq -\frac{1}{aH}(1+\epsilon+\lambda).
\end{equation}
 On the other hand, up to the first order of slow-roll
parameters, Eq. (\ref{ddzk}) can be approximated by
\begin{equation}\label{ddzk40}
\frac{z_k''}{z_k}=2a^2H^2\left(1+\frac{5}{2}\epsilon-\frac{3}{2}\eta-\lambda\right).
\end{equation}
Using Eqs.(\ref{tau39}) and (\ref{ddzk40}), we rewrite the
equation of motion for scalar fluctuation mode (\ref{uk''}) as
\begin{equation}\label{uk''2}
u_k''+\left[c_s^2k^2-\frac{\mu^2-\frac{1}{4}}{\tilde{\tau}^2}\right]u_k=0,
\end{equation}
where the parameter
\begin{equation}
\mu\simeq \frac{3}{2}+3\epsilon-\eta+\frac{2}{3}\lambda.
\end{equation}

These modes are normalized so that they satisfy the Wronskian
condition
\begin{equation}
u_k^*\frac{du_k}{d\tilde{\tau}}-u_k\frac{du_k^*}{d\tilde{\tau}}=-i.
\end{equation}

 On the subhorizon scale $c_s^2k^2\gg z_k''/z_k$, the equation (\ref{uk''2}) has a plane
 wave solution
\begin{equation}\label{solution1}
 u_k=\frac{1}{\sqrt{2c_sk}}e^{-ic_sk\tilde{\tau}}.
\end{equation}
which indicates that perturbation with wavelength within the
horizon oscillate like in flat spacetime. This does not come as a
surprise, since in the UR regime, one expects that approximating
the spacetime as flat is a good approximation.

Taking the solution (\ref{solution1}) as the initial condition, we
can obtain the solution of (\ref{uk''2}) on the superhorizon,
$c_s^2k^2\ll z_k''/z_k$,
\begin{eqnarray}
u_k&\simeq&
\frac{1}{\sqrt{2c_sk}}(-c_sk\tilde{\tau})^{\frac{1}{2}-\mu}\nonumber\\
&\simeq&\frac{1}{\sqrt{2c_sk}}\left[\frac{c_sk(1+\epsilon+\lambda)}{aH}\right]^{\frac{1}{2}-\mu}.
\end{eqnarray}

Thus, we can express the power spectrum on superhorizon scales of
the comoving curvature as
\begin{eqnarray}
P_R(k)&=&\frac{k^3}{2\pi^2}\left|\frac{u_k}{z_k(\tilde{\tau})}\right|^2\nonumber\\
&\simeq&\frac{1}{2\epsilon}\frac{1}{M_{pl}^2}\left(\frac{H}{2\pi}\right)^{2}
\left(\frac{c_sk}{aH}\right)^{3-2\mu}\left(1+\lambda\right)^{-1-2\mu}.
\end{eqnarray}
When the perturbation mode $k$ crosses the Hubble radius,
\begin{equation}
c_s^2k^2=\frac{z_k''}{z_k}=2a^2H^2\left(1+\frac{5}{2}\epsilon-\frac{3}{2}\eta-\lambda\right).
\end{equation}
At the same time, the power spectrum is reduced to be
\begin{eqnarray}
P_R(k)&\simeq&\frac{1}{2\epsilon}\frac{1}{M_{pl}^2}\left(\frac{H}{2\pi}\right)^{2}
\left(2+5\epsilon-3\eta-2\lambda\right)^{-3\epsilon+\eta-\frac{2}{3}\lambda}
\nonumber\\
&\times&\left(1+\lambda\right)^{-4-6\epsilon+2\eta-\frac{4}{3}\lambda}.
\end{eqnarray}
Up to the first order of slow-roll parameters and $\lambda$, we
obtain
\begin{equation}
\frac{d\ln k}{dt}\simeq(1-\epsilon+4\epsilon\lambda)H,
\end{equation}
and
\begin{equation}
\frac{d\lambda}{d\ln k}\simeq-4\epsilon\lambda.
\end{equation}
Therefore, the spectra index of the scalar metric perturbation and
its running can be expressed respectively as follows:
\begin{equation}
n_s-1\equiv \frac{d\ln P_R}{d\ln
k}\simeq-6\epsilon+2\eta+\frac{8(6+\ln 2)}{3}\epsilon\lambda,
\end{equation}
\begin{equation}
\frac{dn_s}{d\ln
k}=14\epsilon\eta-24\epsilon^2-2\xi^2+\frac{16}{3}(6+\ln
2)\epsilon\eta\lambda.
\end{equation}
Obviously, when the parameter $\lambda\rightarrow 0$, the
contribution from the non-commutativity of spacetime to the
spectral index and its running will also vanish. Similar to the
case in the ordinary noncommutative inflation \cite{Huang3}, the
effects of the non-commutativity of spacetime suppress the power
spectrum of the primordial perturbations which lead to a more blue
spectrum with a correction $\frac{8(6+\ln 2)}{3}\epsilon\lambda$
to the spectrum index.

\section{Discussion}
Just as the ordinary inflation scenario, besides the scalar
perturbations that couple to the matter distribution in the
universe and form the "seeds" of the large scale structure,
tachyon inflation both in commutative and non-commutative
spacetime also predicts a tensor perturbation, or called
gravitational perturbation, of the metric. Because the tensor
perturbations during the period of inflation depend only on the
energy scale of the inflation, we can consider that they only
describe the propagation of gravitational waves and do not couple
to the matter term. Therefore, the expressions that describe  the
same as that in ordinary scalar field inflation.

Steer and Vernizzi \cite{Steer} have investigated typical
inflationary tachyon potentials, such as the inverse cosh
potential, the exponential potential and the inverse power law
potential. They also discussed their observational consequences
and compared them with WMAP data. The regime of blue scalar
spectral index and large gravitational waves is not explored by
these potentials. However, the effects of the non-commutativity of
spacetime suppress the power spectrum of the primordial
perturbations which leads to a more blue spectrum for the tachyon
inflation scenario.

\vspace{0.8cm} \noindent ACKNOWLEDGEMENT: This work was partially
supported by NKBRSF under Grant No. 1999075406.

\end{document}